# Elastic colloidal monopoles and reconfigurable self-assembly in liquid crystals

Ye Yuan[1], Qingkun Liu[1], Bohdan Senyuk[1] and Ivan I. Smalyukh[1,2,3*]

[1]Department of Physics, University of Colorado, Boulder, CO 80309, USA

[2]Department of Electrical, Computer, and Energy Engineering, Materials Science and Engineering Program and Soft Materials Research Center, University of Colorado, Boulder, CO 80309, USA

[3]Renewable and Sustainable Energy Institute, National Renewable Energy Laboratory and University of Colorado, Boulder, CO 80309, USA

* Correspondence to: ivan.smalyukh@colorado.edu

Monopole-like electrostatic interactions are ubiquitous in biology[1] and condensed matter[2–4], but they are often screened by counter-ions and cannot be switched from attractive to repulsive. In colloidal science, where the prime goal is to develop colloidal particles[2,3] that mimic and exceed the diversity and length-scales of atomic and molecular assembly, electrostatically charged particles cannot change the sign of their surface charge or transform from monopoles to higher-order multipoles[4]. In liquid-crystal colloids[5–7], elastic interactions between particles arise to minimize the free energy associated with elastic distortions in the long-range alignment of rod-like molecules around the particles[5]. In dipolar[6,8], quadrupolar[8–12] and hexadecapolar[13] nematic colloids, the symmetries of such elastic distortions mimic both electrostatic multipoles[14] and the outmost occupied electron shells of atoms[7,15,16]. Electric and magnetic switching[17,18], spontaneous transformations[19] and optical control[20] of elastic multipoles, as well as their interactions with topological defects and surface boundary conditions, have been demonstrated in such colloids[21–23]. However, it has long been understood[5,24] that elastic monopoles should relax to uniform or higher-order

**multipole states because of the elastic torques that they induce[5,7]. Here we develop nematic colloids with strong elastic monopole moments and with elastic torques balanced by optical torques exerted by ambient light. We demonstrate the monopole-to-quadrupole reconfiguration of these colloidal particles by unstructured light, which resembles the driving of atoms between the ground state and various excited states. We show that the sign of the elastic monopoles can be switched, and that like-charged monopoles attract whereas oppositely charged ones repel, unlike in electrostatics[14]. We also demonstrate the out-of-equilibrium dynamic assembly of these colloidal particles. This diverse and surprising behaviour is explained using a model that considers the balance of optical and elastic torques that are responsible for the excited-state elastic monopoles and may lead to light-powered active-matter systems and self-assembled nanomachines.**

Microfabricated thin silica particles with shapes of hexagonal prisms were surface-functionalized with azobenzene dye molecules to form molecular surface monolayers[25-27] (Extended Data Fig. 1). When dispersed in a nematic liquid crystal (LC)[4], such particles distort the alignment field of rod-like molecules, the nonpolar director field $\mathbf{n}(\mathbf{r})\equiv-\mathbf{n}(\mathbf{r})$ with the head-tail symmetry[5], due to the anisotropic interaction of the LC molecules with the colloidal particle's surfaces[7] (Fig 1). The particle-induced elastic distortions manifest themselves as spatial patterns of transmitted light intensity and color when the LC colloidal sample is placed between polarizers and observed in an optical microscope using monochromatic (Fig. 1a,b) or white (Fig. 1i,j,and l) light. Unlike conventional LC colloids, for which control of $\mathbf{n}(\mathbf{r})$-structures around particles requires high-power focused laser beams of optical tweezers[20], our photo-responsive nematic colloids exhibit polarizing optical microscopy textures and director structures highly dependent on intensity, color and polarization of low-intensity light, including microscope and ambient light (Fig. 1). Such facile reconfiguration of **n(r)**-structures emerges from the collective reorientation of the azobenzene moieties[25-27] within monolayers on the surface of particles when exposed to light, which alters the molecular interactions at LC-colloidal interfaces, often causing rotation of particles and twist of **n(r)** around them with respect to the uniform far-field director **n**$_0$.

When observed with a conventional optical microscope, the particles appear to have a vector connecting two diametrically opposite vertices of the colloidal platelet's hexagonal facet rotated with respect to **n**$_0$ to an angle $|\theta|$<30º. For each particle, the clockwise versus counterclockwise rotation is selected randomly for unpolarized light or light with linear polarization along **n**$_0$ (Fig. 1a-k) due to spontaneous symmetry breaking, but it can be controllably chosen by slightly rotating the linear polarization direction away from **n**$_0$. Rotated to the maximum angle of $\theta$ above and below the platelet, **n(r)** smoothly twists back to the uniform alignment along **n**$_0$ far from the particle surfaces (Fig. 1g,h and Extended Data Fig. 1). Particles with opposite sense of **n(r)** rotation have different appearance when observed in both monochromatic (Fig. 1a,b) and white light (Fig. 1i,j) between polarizers with or without an additional phase retardation plate in the optical pathway. To avoid the spontaneous platelet rotation away from $\theta$=0º, we can use red imaging light while carefully shielding the sample from ambient white light or by aligning linear polarization of blue or achromatic light orthogonally to **n**$_0$ (Fig. 1l,m and Extended Data Figs. 2,3). This spectrally selective and polarization-dependent sensitivity of our colloids to light is consistent with the response of the azobenzene dye molecules self-assembled on particle surfaces[25-27].

Perturbations of **n(r)** away from **n**$_0$ far from the rotated particles are monopole-like at $\theta\neq$0º (Fig. 1e,f,k), but become of quadrupolar type at $\theta$=0º (Fig. 1m). The twisted monopole-type director structure induced by a rotated platelet has low symmetry (Fig. 1g,k). Differently, the quadrupolar structure has 3 mirror symmetry planes (Fig. 1l,m), one of which is orthogonal to both the platelet and $n_0$ and two are parallel to $n_0$, with one also parallel to the platelet and one orthogonal to it. By numerically minimizing the elastic energy due to $n(r)$-distortions (see Methods) that are created in response to the synchronous rotation of the thin platelet and the director at its surface within a large LC sample, we obtain $n(r)$-configuration shown in Fig. 1h and Extended Data Fig. 4. The multipole expansion analysis[15] for this $n(r)$ yields only the strongly pronounced monopole moment, with higher-order multipole moments equal zero up to numerical precision. The elastic charge of the monopole has the same sign as $\theta$ and the strength of the monopole moment increases[8-10] with $|\theta|$ (see Methods).

Using laser tweezers and video microscopy, we probe colloidal pair interactions (Fig. 2 and Extended Data Fig. 5) by releasing particles with elastic charges of different strength and signs from the laser traps near each other and at different center-to-center separation vector $r_c$ orientations relative to $n_0$. Unlike in electrostatics, the like-charged elastic monopoles with the same signs of $\theta$ attract (Fig. 2a-c and Extended Data Fig. 5), but oppositely charged ones repel (Fig. 2d-f and Extended Data Fig. 5). This is consistent with the LC's tendency to minimize the distortions of $n(r)$ deviating away from $n_0$ around the particles (Fig. 3a,b), where like-charged elastic monopoles with the same sense of particle rotation and sign of $\theta$ can reduce these elastic distortions when spatially co-located (Fig. 3b) whereas opposite-charged elastic monopoles can only minimize the free energy by increasing pair separation (Fig. 3a). Since our colloidal system is highly overdamped (Reynolds number <<1), the inertia effects are negligible and the elastic interaction forces are balanced by the viscous drag forces[11] $f_d \propto dr_c/dt$. For a $2^l$ elastic multipole (here the integer $l$ determines the order of a multipole), this balance with the elastic force $f_e \propto 1/r_c^{2l+2}$ results in a differential equation $dr_c/dt = -\alpha/r_c^{2l+2}$, which yields the anticipated time dependence of inter-particle distance $r_c(t)=[r_0^{2l+3}-(2l+3)\alpha t]^{1/(2l+3)}$, where $r_0$ is the initial center-to-center distance at the time of releasing particles from tweezers and $\alpha$ is a fitting parameter dependent on the particle's viscous drag and strength of multipole moments. For monopoles $l=0$ and $r_c(t)=(r_0^3-3\alpha t)^{1/3}$. Experimental data for $r_c(t)$, interaction potential and force versus distance are consistent with the elastic monopole nature of the rotated particle (Fig. 2 and Extended Data Fig. 5). As expected, elastic monopoles exhibit pair potential $\propto \pm 1/r_c$ (Fig. 2c,f and Extended Data Fig. 5). While the particles interact, the sign of their monopole moments can be switched by simply rotating the linear polarization of light in an opposite direction (Fig. 3a-c and Supplementary Video 1). Consequently, when this is done for one of the two interacting particles, the interactions are also switched from repulsion to attraction and vice versa (Fig. 3c). In addition, the Brownian motion of such a monopole platelet exhibits very different angular dependencies of anisotropic diffusion relative to $\mathbf{n}_0$ compared to that of the platelet in the "quadrupolar state" (Fig. 3d,e) because of the different symmetries of $\mathbf{n}(\mathbf{r})$ around monopolar and quadrupolar colloidal particles as well as the particle orientation.

Remarkably, the intensity of the blue light needed to induce the strong monopole moments and switch their signs is very low, with the power <1 nW per particle (Fig. 3), as estimated by measuring the power of light uniformly illuminating the sample within the platelet's area[25]. At slightly higher powers (>5 nW) of light shining on the pair of interacting particles[25], they start periodic rotational spinning while continuing to undergo elastic interactions mediated by the LC host medium (Fig. 4a-c and Supplementary Video 2). One can understand such periodic particle rotations as a result of an effective feedback mechanism that emerges from the rotation of azobenzene moieties and particle in response to polarized light. This particle reotation twists $\mathbf{n}(\mathbf{r})$ and consequently changes polarization of light traversing through the LC-embedded particle, which in turn further rotates the particle and so on[25]. This out-of-equilibrium behavior of a pair of colloidal objects leads to their dynamic self-assembly, where individual particles spin before approaching each other and after forming a colloidal dimer (Fig. 4a). The elastic interaction accompanied by spinning does not resemble interaction of monopoles, but rather is consistent with the interaction of elastic quadrupoles (Fig. 4 and Extended Data Fig. 6), with $r_c(t)=(r_0^7-7\alpha t)^{1/7}$ and interaction potential $\propto -1/r_c^5$. The spinning frequencies of the two particles are different and their $\theta$ (and also the corresponding monopole moments) change between negative and positive values in an uncorrelated way and on timescales much shorter than the time of the long-ranged interaction (Fig. 4b,c). Therefore, the instantaneous elastic monopole moments periodically switch from being like-charged to opposite-charged and, on average, do not contribute to the observed behavior (Fig. 4b). The ensuing scaling of pair interaction potential $\propto -1/r_c^5$ is similar to that obtained for linear polarization of excitation/imaging light orthogonal to $\mathbf{n}_0$ (Fig. 4d,e and Extended Data Fig. 6) and/or when only red but not violet/blue light is used, corresponding to quadrupolar nematic colloids (Fig. 1l,m).

Light with linear polarization $\mathbf{P}_e \parallel \mathbf{n}_0$ traverses through the LC as an extraordinary eigen mode (Fig. 1a,b). As it reaches the platelet with the azobenzene molecular monolayers, it torques the trans-state moieties of the dye molecules to rotate and align orthogonally to[25,27] $\mathbf{P}_e \parallel \mathbf{n}_0$, driven by the azobenzene's tendency to avoid the excited state[28]. Due to the mechanical coupling between the azobenzene monolayer and $\mathbf{n}(\mathbf{r})$ through the surface anchoring, a platelet is prompted to rotate, albeit this rotation further increases $\mathbf{n}(\mathbf{r})$-distortions around the platelet (Fig. 1). The optical torque, transferred to the platelet through the surface anchoring, is then balanced by the elastic torque tending to minimize the free energy of elastic distortions (Fig. 1g and Extended Data Fig. 1). In addition to numerical modeling (Fig. 1h and Extended Data Fig. 4), approximating the platelet by a high-aspect-ratio disk of radius $R$, the strength of elastic torque due to the director locally rotated by the platelet to an angle $\theta$ can be estimated using theory of Brochard and de Gennes[24] as $T_e=8KR\theta$, where $K$ is the average elastic constant of the LC. Consistent with our numerical modeling (Extended Data Fig. 7) and experiments (Fig. 1c,d), the rotated platelet is accompanied by twisted $\mathbf{n}(\mathbf{r})$ around it whereas adiabatic rotation of light's nearly linear polarization closely follows this twisted $\mathbf{n}(\mathbf{r})$ above and below the platelet, albeit slightly lagging it[29]. Under these conditions, for independently measured material and geometric parameters (see Methods), competition of optical and elastic torques yields the torque-balancing angle $\theta_b \approx 26°$ that agrees with experiments (Fig. 1), as well as the excited-state elastic monopoles.

Although our monopoles and samples are three-dimensional in nature, to facilitate direct optical microscopy observations of colloidal behavior, the geometry of our experiments is designed as quasi-two-dimensional because elastic monopoles and quadrupoles repel from the confining cell substrates with strong surface boundary conditions[20] for the director along $\mathbf{n}_0$. These boundary conditions and confinement also tend to align the large-area faces of platelets to be parallel to the confining substrates, facilitating their self-assembly within the sample midplane. The light illumination direction is orthogonal to both the platelets and confining substrates, defining the axis of platelet rotation that distorts $\mathbf{n}(\mathbf{r})$. Without the influence of confining plates, the interactions between platelets and other anisotropic colloidal objects should depend on their relative orientations in a manner that has more structure than a simple scalar elastic charge. It will be of interest to extend and generalize our study to LC colloids without such confinement (which also causes screening of elastic interactions at large inter-particle distances) and with different geometric shapes because the combination of LC and particle anisotropy may lead to a richer behavior exceeding that of the electrostatic analogs of our colloids.

To conclude, we have designed, demonstrated and explained the elastic colloidal monopoles in LCs that can be induced by ambient-intensity light, with switchable sign and amplitude of the monopole moment. We have shown how like-charged elastic monopoles attract and opposite-charged ones repel, as well as how these interactions can be switched from attractive to repulsive and even how monopoles can be transformed to quadrupoles by invoking the highly reconfigurable nature of our LC colloids. Our work is a contribution towards the development of the soft matter analogs of Rydberg matter[16], where colloidal atoms can self-assemble while in out-of-equilibrium states into structures different from what they would form under equilibrium conditions, though at length and energy scales very different from those in the atomic Rydberg matter. This may allow for out-of-equilibrium excited-state self-assembly of composite materials based on LCs and may lead to new breeds of active matter[30].

**Data availability:** All data generated or analyzed during this study are included in the published article and its Supplementary Information and are available from the corresponding author on a reasonable request.

**Code availability:** The codes used in this study for the numerical simulation and calculation are available upon request.

**Acknowledgments:** We thank P. Ackerman, A. Hess, S. Park, B. Tai and M. Tasinkevych for technical assistance and discussions.

**Funding:** This research was supported by the U.S. Department of Energy, Office of Basic Energy Sciences, Division of Materials Sciences and Engineering, under Award ER46921, contract DE-SC0019293 with the University of Colorado at Boulder.

**Author Contributions:** Y.Y. and Q.L. synthesized azobenzene-containing dye molecules. Q.L. fabricated colloidal particles. Y.Y. and B.S. performed experiments and Y.Y. performed numerical modeling. Y.Y., B.S. and I.I.S. analyzed data. I.I.S conceived and directed the project, designed experiments, provided funding and wrote the manuscript, with the input from all authors.

**Competing interests:** Authors have no competing interest.

**Methods**

**Sample preparation**. The photo-responsive azobenzene molecules were synthesized by combining 2.70 g of methyl red, 2.22 g of 1,3-dicyclohexylcarbodiimide, and 2.38 mL of (3-aminopropyl) triethoxysilane in 60 mL of dichloromethane under nitrogen[27]. The resulting solution was stirred overnight and filtered, after which column chromatography (50% ethyl acetate, 50% hexane in volume) and rotary evaporation were used to isolate and purify the obtained azobenzene-containing dye molecules. To fabricate hexagonal silica micro-platelets, a 0.5 µm thick silica layer was first deposited on a silicon wafer by plasma-enhanced chemical vapor deposition and then an additional layer of photoresist AZ5214 was obtained by spin-coating. A photomask was then produced on the photoresist layer by a direct-laser writing system (DWL 66FS, Heidelberg Instruments) to define particle shapes. Finally, free-standing silica platelets were carved out of the substrate by inductively coupled plasma etching after removal of the photoresist by acetone. The photosensitive azobenzene molecular monolayers were self-assembled on hydroxylated surfaces of the platelets. To achieve this, the micro-platelets, while still on the substrate, were first treated in piranha solution (98 wt% sulfuric acid and 30 wt% hydrogen peroxide, volume ratio 1:1) for 1 h, rinsed thoroughly with deionized water and dried at 70 °C for 1 h. Subsequently, the platelets were immersed in 20 mL of 0.53 mM toluene solution of the azobenzene molecules, followed by injecting 20 µL of n-butylamine. After reacting overnight at 45 °C, the platelets were rinsed with toluene and cured at 120 °C for 2 hrs. Isopropanol suspension of the functionalized platelets was obtained by ultrasonication to release them from the substrate and was then mixed with a nematic LC. The mixture was left at room temperature uncovered to allow the solvent to evaporate, eventually yielding LC colloidal dispersion. The resulting dispersion was then sandwiched between glass plates separated by glass spacers with diameter within 4-16 µm and sealed with epoxy. Prior to this, thin layers of polyimide (PI2555, from HD Microsystem) were spin-coated on the glass plates and rubbed unidirectionally to define planar boundary conditions and orient the far-field director $\mathbf{n}_0$ along the rubbing direction. All chemicals used were obtained from Sigma-Aldrich unless noted otherwise.

**Details of the experimental imaging, laser trapping and illumination setup.** Polarizing optical microscopy and video microscopy of the colloidal interactions between the platelets were performed using a home-made micro-projection system built based on an upright microscope (BX51, Olympus) and LC microdisplays extracted from a consumer color projector (EMP-730, Epson) (Extended Data Fig. 2). Since a conventional white-light microscope observation prompts rotation of platelets and induces monopoles, we additionally studied our nematic colloids with light narrow-band optical microscope illumination specially designed to preclude its interaction with the azobenzene monolayers and particles. Due to azobenzene's sensitivity to violet-blue light (Fig. 3a), red light provided by filtering microscope illumination with a bandpass filter (FF01-640/14-25, Semrock) was used for imaging and observation in this case. After passing through the sample, the imaging light was collected by high numerical aperture objective lenses with magnification 20× to 60× (Olympus) and directed to a charge-coupled device (CCD) camera (Grasshopper3, PointGrey) to record images and videos. The blue excitation light that controllably rotates the azobenzene molecules and twists the director field was produced by projecting patterns designed in Microsoft PowerPoint slides and projected using the LC microdisplays[31] (Extended Data Fig. 2). Polarization of the projected blue light was controlled by rotatable linear polarizers inserted within the optical path of the two illumination channels (Extended Data Fig. 2). Generation and switching of elastic monopoles with opposite signs requires the ability of a local control of the polarization state of blue light projected within one field of view of a microscope on scales comparable to particle dimensions. In our experiments, this is achieved using a home-built setup shown in Extended Data Fig. 2. The mutually orthogonal linear polarization of the blue light from the two separately controlled channels of the setup allowed separating them apart by a polarizing beam splitter cube (CCM1-PB251, obtained from Thorlabs) after exiting the projector. Additional retardation plates and linear polarizers inserted into the two separate light paths enabled independent control of the polarization within the two patterns separately generated by two microdisplays. Light rays from the two separate channels were then recombined with a 50/50 plate beamsplitter (BSW10R, from Thorlabs) before entering the microscope and projected to the sample placed on a microscope stage within the same field of view of an objective. Using this setup (Extended Data Fig. 2), we illuminated sample areas with blue-light patterns having controlled linear polarization states. In particular, this allowed for illuminating particles in proximity of each other using blue light with linear polarization states independently controlled at will. Integrated with the illumination and imaging setup, laser tweezers operating at 1064 nm and 1~100 mW (YLR-10-1064, IPG Photonics) allowed us to set proper initial conditions for the platelets suspended in LC in order to probe colloidal pair interactions. The trajectories of individual particles in the videos taken under microscope were extracted using a freeware ImageJ (obtained from the National Institute of Health). Details of the laser tweezers setup and operation are described in Ref. 20.

**Characterization of colloidal diffusion and pair interactions**. Intrinsic anisotropy of the LC host, anisotropic shape of studied particles and formation of defects and director distortions all contribute to the anisotropic Brownian motion of the colloidal platelet (Fig. 3d,e). Interestingly, this anisotropic Brownian motion can be further altered by tuning the polarization of the blue excitation light, as characterized by the angular dependence of the diffusion constant (Fig. 3d,e). The diffusion constants $D$ are calculated based on the single particle Brownian motion statistics by fitting the displacement histogram with a Gaussian function[20,32], yielding Stokes' drag coefficients at different directions relative to $\mathbf{n}_0$ according to $c=k_B T/D$, where $T=300$ K is temperature and $k_B=1.38\times 10^{-23}$ J K$^{-1}$ is the Boltzmann constant. For simplicity, an average value of $c=10^{-6}$ kg s$^{-1}$ was often used in the analysis of forces and potentials. The viscous drag force was calculated based on the speed of the colloidal platelet $v=dr_c/dt$ according to $f_d=cv$ and was balanced by the elastic force between the interacting particles (because of the highly overdamped motion, as discussed in the main text). This allowed for determining the distance dependence of the elastic colloidal interaction potential and force, which both were found to be in agreement with theoretical expectations based on electrostatic analogy (Figs. 2, 4 and Extended Data Fig. 5).

The monopole-type nature was revealed and the strength of the elastic charges was estimated by fitting the experimental data using the power law functions of monopole-monopole interaction. Similar to that of electrostatics[14], the force between two elastic charges of strength $Q$ reads[9,10]

$$f_e = \pm 4\pi K \frac{Q^2}{r_c^2}, \qquad (1)$$

where $K$ is the average elastic constant of the LC and the sign depends on the relative sign of the elastic charges. One can define a parameter

$$\alpha = \pm 4\pi K Q^2/c \qquad (2)$$

so that the force balance equation $f_d+f_e=0$ is reduced to

$$\frac{dr_c}{dt} = -\frac{\alpha}{r_c^2}, \qquad (3)$$

which is integrated to give the equation of motion:

$$r_c(t) = (r_0^3 - 3\alpha t)^{1/3}, \qquad (4)$$

where $r_0$ is the initial separation upon releasing the interacting particles from the laser traps. Using the average elastic constant of 4'-$n$-pentyl-4-cyanobiphenyl (5CB) $K$=6.5 pN and value of $\alpha$ obtained from fitting experimental data of separation distance versus time, one obtains the magnitude of the corresponding elastic charge

$$Q = \sqrt{|\alpha|c/4\pi K}, \qquad (5)$$

which is found to be ~0.5 µm in our experiments (Fig. 2 and Extended Data Fig. 5).

On the other hand, elastic charges, or, in general, the strength of multipole moments, can be calculated directly using the known far-field director configurations. To do this, we consider a spherical surface enclosing the platelet with radius $a$>>$R$, where $R$ is the effective lateral size of the platelet, so that director deviation away from $\mathbf{n}_0$=(0, 0, 1) is small. Assuming the director orientation on this surface is uniform $\mathbf{n}_S$=(sin $\theta_S$, 0, cos $\theta_S$), the total charge is given by[10]

$$Q = \frac{a}{4\pi} \int dS \, \frac{\sin \theta_S}{a^2} = a \sin \theta_S \approx a\theta_S, \qquad (6)$$

where $\theta_S$ is small and cos $\theta_S$~1. At a distance $a$=10 µm, the director orientation angle is $\theta_S$=3°, which is closely related to how strongly the platelet perturb $\mathbf{n}(\mathbf{r})$, i.e. the platelet rotation angle $\theta$.

**Polarimetric imaging and analysis.** The birefringence of LC and distortion of the director due to colloidal inclusions cause changes in the polarization state of the light traversing the sample[29,33]. These changes are monitored by polarimetric imaging of the exiting light in the experiments when using linearly polarized incoming light. To perform polarimetric imaging, a quarter-wave plate (AQWP05M-600, obtained from Thorlabs) was inserted right below the microscope's analyzer whose polarization was set along the far-field director $\mathbf{n}_0$. Images were taken when the fast axis of the quarter-wave plate was rotated from 0° (defined with respect to $\mathbf{n}_0$) to 180° at steps of 22.5°. These images were processed numerically to calculate the ellipticity and orientation of the outgoing light's polarization ellipses[33]. The image analysis process is briefly summarized as follows. Assuming the angle that the fast axis of the quarter-wave plate was rotated to be $\beta_n$ ($\beta_n = 22.5° \cdot n, n = 1, ...,8$), four coefficients were defined for each pixel of images

$$C_1 = \frac{2}{N}\sum_{n=1}^{8} I_n, \tag{7}$$

$$C_2 = \frac{4}{N}\sum_{n=1}^{8} I_n \sin 2\beta_n, \tag{8}$$

$$C_3 = \frac{4}{N}\sum_{n=1}^{8} I_n \cos 4\beta_n, \tag{9}$$

$$C_4 = \frac{4}{N}\sum_{n=1}^{8} I_n \sin 4\beta_n, \tag{10}$$

from which the Stokes parameters[29] were calculated on the pixel-by-pixel basis as follows:

$$S_0 = C_1 - C_3, \ S_1 = 2C_3, \ S_2 = 2C_4, \ S_3 = C_2. \tag{11}$$

Images captured by the CCD camera under different $\beta_n$ provided values for the intensity $I_n$, and thus allowed for reconstruction of polarization states of the exiting imaging light after passing through the sample on a pixel-by-pixel basis (Fig. 1c,d). For this, the ellipticity $e$ and the polarization ellipse's orientation angle $\chi$ (defined as the angle between the long axis of the polarization ellipse and $\mathbf{n}_0$) were obtained using $\sin 2e = S_3/S_0$ and $\tan 2\chi = S_2/S_1$ on a pixel-by-pixel basis (Fig. 1c,d). The two-dimensional polarimetric images like the ones shown in Fig. 1c,d were formed by overlaying the plots of orientation patterns of the long axis of ellipse (represented by rods) and ellipticity (shown using colors).

**Estimation of platelet's torque-balancing rotation angle.** Disregarding the energetic costs of director distortions near the edge faces of our thin platelets, the elastic energy due to the twist of director around the rotated platelets can be estimated using the model developed by Brochard and de Gennes for small inclusions in LC exerting a torque on the director structure[24], which gives

$$F_e = 2\pi KC\theta^2, \tag{12}$$

where $C$ is the effective particle's "capacitance" (note that this expression was also derived using electrostatic analogies) and $\theta$ is the net maximum director twist angle with respect to $\mathbf{n}_0$, which is also assumed to be the angle to which a platelet rotates upon exposure to the blue light. The capacitance of the thin hexagonal platelets used in this study can be approximated to be that of high-aspect-ratio disks[34], $C=2R/\pi$. Therefore, the elastic torque from the director distortion caused by balancing of the optical torque reads

$$T_e = \frac{\partial F_e}{\partial \theta} = 8KR\theta. \tag{13}$$

The free energy potential describing the coupling between polarized blue light and director orientation through the azobenzene monolayers can be expressed as[25]

$$F_p = \frac{1}{2}\sigma(I,e) \int dS (\mathbf{P}_p \cdot \mathbf{n}_p)^2, \tag{14}$$

where $\sigma(I,e)$ characterizes the strength of the monolayers' surface anchoring and is dependence on the intensity $I$ and ellipticity $e$ of the blue light reaching the surface of the platelet[25]. In the expression above, $\mathbf{P}_p$ is the orientation of the long axis of polarization ellipse of the blue light and $\mathbf{n}_p$ is the orientation of $\mathbf{n}(\mathbf{r})$ at the platelet's surface. Assuming uniform $\mathbf{n}_p$ across the large-area surface of the platelet, the free energy potential is simplified to read

$$F_p = \frac{1}{2}\sigma(I,e) \int dS \cos^2\phi = \frac{1}{2}\sigma(I,e) 2A \cos^2\phi, \tag{15}$$

where $\phi$ is the angle between $\mathbf{P}_p$ and $\mathbf{n}_p$ and $A$ is the area of the hexagonal surface of the platelet. The factor of 2 accounts for the presence of 2 (top and bottom) photoresponsive surfaces interfacing the LC surfaces. Following the procedures described in the Ref. 29, the angle $\phi$ characterizing the close to linear polarization state of light traversing the platelet is calculated using the Jones matrix method

$$\phi = \frac{1}{2}\tan^{-1}\frac{2\theta X \tan X}{(\theta^2-\Gamma^2/4)\tan^2 X-X^2}, \quad (16)$$

where

$$\Gamma = \frac{2\pi\Delta n d}{\lambda}, X = \sqrt{\theta^2+\left(\frac{\Gamma}{2}\right)^2} \quad (17)$$

and where $d=12$ μm is the thickness of the LC that the blue light propagates through before reaching the top surface of the platelet, $\lambda=490$ nm is the wavelength of the used blue light, and $\Delta n=0.2$ is the birefringence of 5CB. The value of $\Gamma$ is found to be 30.8, much larger than that of $\theta$, with both expressed in radians (experimental observation indicates that $\theta$ is usually in the range of 30°~0.52 rad or less, as shown in Fig. 1). In this case, $\phi$ is approximated by taking the first two non-zero terms of the Taylor expansion given for $\lim_{\theta/\Gamma\to 0}\phi\to 0$:

$$\phi \sim -\frac{1}{2}\left(0.91\frac{2\theta}{\Gamma}+0.42\left(\frac{2\theta}{\Gamma}\right)^3\right). \quad (18)$$

The corresponding optical torque transferred to the platelets through anisotropic surface interactions and causing platelet rotation reads:

$$T_{\mathrm{p}} = \frac{\partial F_{\mathrm{p}}}{\partial \phi}|_{\phi\sim 0} \approx -2\sigma A\phi. \quad (19)$$

Balancing the optical torque and the elastic torque under the equilibrium conditions allows for finding the platelet rotation angle analytically while assuming $A=\pi R^2$:

$$\theta_{\mathrm{tb}} = \frac{\Gamma}{2}\sqrt{\left(\frac{4K\Gamma}{\pi\sigma R}-0.91\right)/0.42}. \quad (20)$$

Using the average elastic constant of 5CB $K=6.5$ pN, $R=2.8$ μm and $\sigma=10^{-4}$ J m$^{-2}$, we obtain the experiment-matching torque-balancing angle $\theta_{\mathrm{tb}}=26°$, where the surface anchoring strength $\sigma$ at ~1 nW light intensity was estimated independently.

**Numerical modeling of director configurations and polarization states.** Computer simulations of the director distortion induced by the azobenzene-capped platelets were based on minimization of Frank-Oseen free energy[5]

$$F = \int dV \left[ \frac{1}{2} K_{11} (\nabla \cdot \mathbf{n})^2 + \frac{1}{2} K_{22} (\mathbf{n} \cdot \nabla \times \mathbf{n})^2 + \frac{1}{2} K_{33} (\mathbf{n} \times \nabla \times \mathbf{n})^2 - K_{24} (\nabla \cdot (\mathbf{n}(\nabla \cdot \mathbf{n}) + \mathbf{n} \times (\nabla \times \mathbf{n}))) \right], \quad (21)$$

where the Frank elastic constants $K_{11}$, $K_{22}$, $K_{33}$ and $K_{24}$ describe the energetic costs of the splay, twist, bend and saddle-splay deformations of the director field $\mathbf{n}(\mathbf{r})$. This modeling was done under the assumption of infinitely thin platelets with fixed, uniform in-plane boundary conditions, with the platelets rotated to $\theta_{tb}$ with respect to their orientation under no illumination light conditions. Following Ref. 35, a large grid size of 101×101×101 was used with grid spacing of $h_x=h_y=h_z=0.5$ μm. The far-field director was set to be $\mathbf{n}_0=(0,0,1)$ with fixed boundary conditions on the top and bottom (xz plane) and periodic boundary conditions in the lateral directions (xy and yz plane) of the computational box. At the central grid point with coordinates (51,51,51), the orientation of the director was fixed at an angle $\theta_\mathbf{n}=30°$ or $45°$ with respect to $\mathbf{n}_0$, i.e. $\mathbf{n}=(\sin\theta_\mathbf{n}, 0, \cos\theta_\mathbf{n})$, while the director at the surrounding grid points were allowed to re-orient accordingly during the simulation. Relaxation of this initial condition yielded director perturbation that decayed as $1/r$ (Extended Data Fig. 4) and monopole moment being the only non-zero multipole moments up to the computational precision.

Propagation of light through such distorted director structures was modeled using the Jones matrix method[29]. To better model the real experimental system, where gravity shifts the platelet downwards with respect to the LC cell midplane, the director region with strong twist was placed closer to the bottom of the computational box, with the director field across the sample thickness parametrized as $\mathbf{n}=(\sin\theta_\mathbf{n}, 0, \cos\theta_\mathbf{n})$. After obtaining the director structure using the approach described above, Jones matrix method was applied assuming each grid point represented a thin layer of 0.16-μm-thick LC with uniform optical axis defined by the local director orientation **n(r)**. This grid spacing of 0.16 μm resulted in total cell thickness of 16 μm as used in the experiments and initial twist at about 3 μm above the bottom substrate. Then the polarization states of the red imaging light ($\lambda$=640 nm) were calculated on a point-by-point basis by propagating the light sequentially through the layers in $y$ direction. Given incident polarization $\mathbf{P}_i \parallel \mathbf{n}_0$, distribution of the ellipticity $e$ and orientation angle $\chi$ of the polarization ellipses were obtained, which corresponds to the change of polarization states as the director perturbation decays away from the edge of the particle (Extended Data Fig. 7). These results are consistent with our polarimetric measurements (Fig. 1c, d).

**Additional References**

# Figures

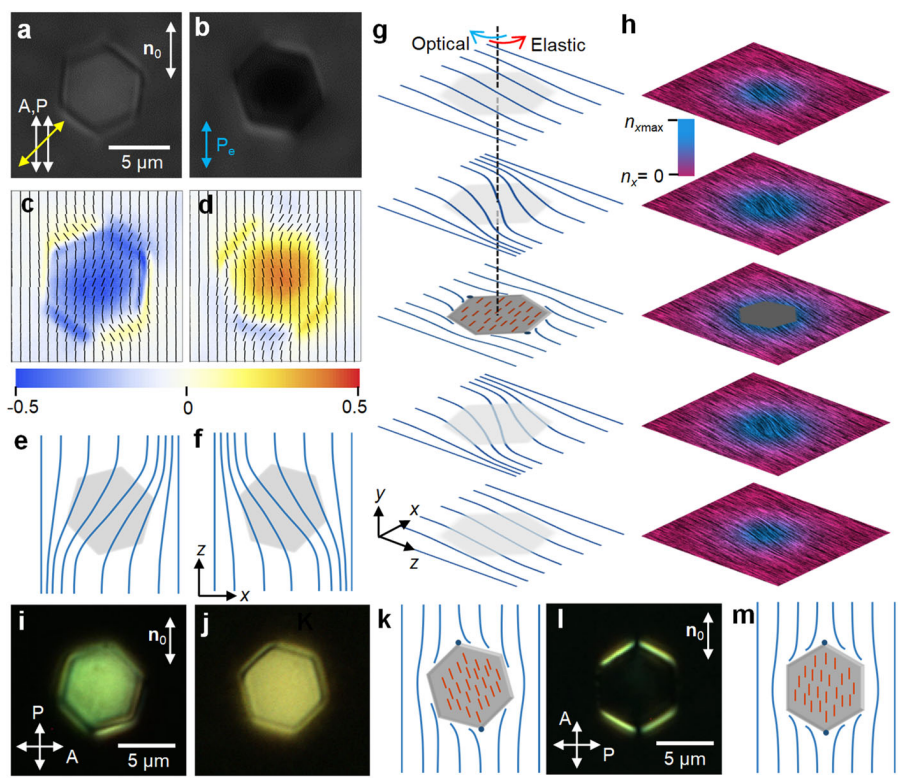

**Fig. 1 | Oppositely charged elastic colloidal monopoles. a,b**, Red-light polarizing optical micrographs of platelets with azobenzene surface monolayers in a nematic LC under excitation light with $\mathbf{P}_e \| \mathbf{n}_0$. "P" and "A" mark the linear polarization directions (white double arrows) of the polarizer and analyzer of the microscope. A yellow double arrow shows the fast axis of a broadband quarter-wave plate. **c,d**, Patterns of polarization state of red imaging light after passing through the sample. Black rods indicate the long axes of polarization ellipses. Colors represent ellipticity (see color bar). **e,f**, $\mathbf{n}(\mathbf{r})$ (blue lines) above platelets for monopoles of opposite signs. (**c**) and (**e**) correspond to (**a**). (**d**) and (**f**) correspond to (**b**). **g**, $\mathbf{n}(\mathbf{r})$ around the platelet; red rods show trans-state azobenzene molecules within the monolayers. Blue and red arrows indicate the optical and elastic torques, respectively. **h**, Numerically simulated pattern of the *x*-component of $\mathbf{n}(\mathbf{r})$ corresponding to (**g**). **i,j**, White-light polarizing micrographs at different platelet orientations for $\mathbf{P}_e \| \mathbf{n}_0$. **k**, $\mathbf{n}(\mathbf{r})$ and azobenzene orientations corresponding to (**j**). $\mathbf{n}(\mathbf{r})$ for (**i**) is a mirror image of (**k**). **l**, Micrograph of a platelet in equilibrium state under white light with $\mathbf{P}_e \perp \mathbf{n}_0$. **m**, Corresponding $\mathbf{n}(\mathbf{r})$ and azobenzene orientations. Filled circles at the vertices of platelets in (**g,k** and **m**) show surface point defects.

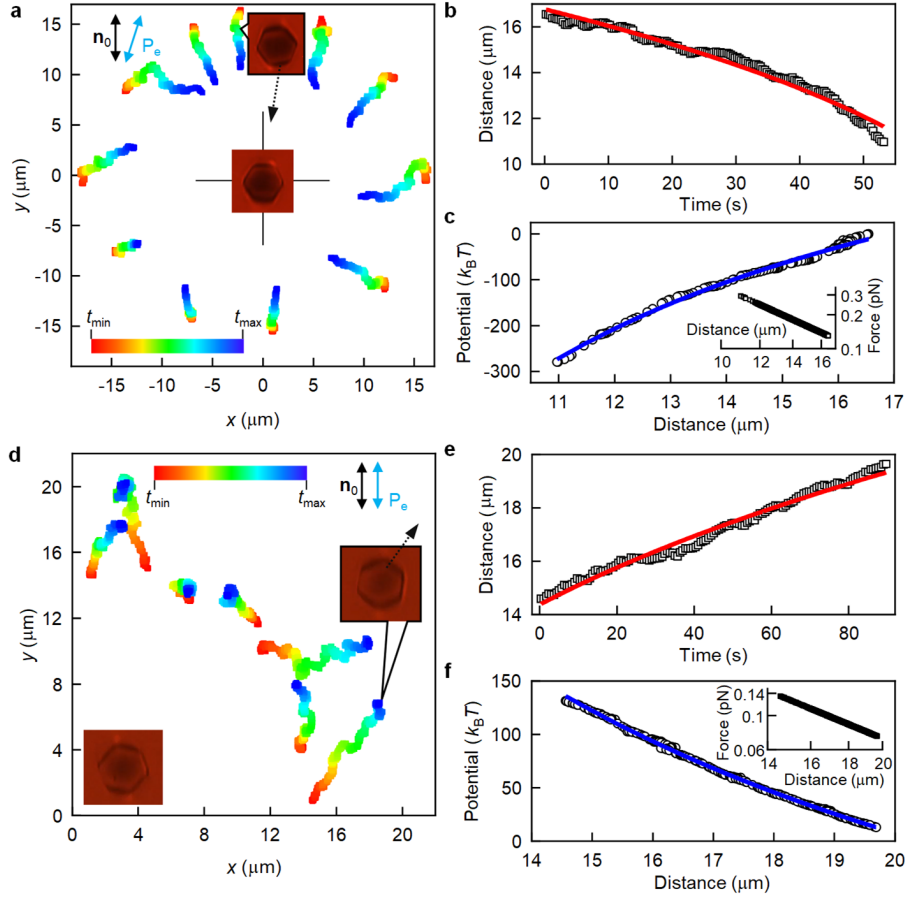

**Fig. 2 | Colloidal interactions between elastic monopoles. a**, Time-color-coded trajectories of platelets after they are released from laser traps, showing attraction between monopoles of the same sign at all orientations of $\mathbf{r}_c$ relative to $\mathbf{n}_0$ within time periods $t_{max}$-$t_{min}$~50-100 s. Polarization of excitation light $\mathbf{P}_e$ is at ~20° with respect to $\mathbf{n}_0$. Insets are micrographs of interacting particles; black arrow indicates direction of relative motion. **b**, Separation distance versus time for particles attracting along $\mathbf{n}_0$. **c**, Interaction potential versus distance corresponding to (**b**), with inset showing a log-log plot of force versus distance. **d**, Omnidirectional repulsion between two monopoles of opposite signs shown within one quadrant; time periods $t_{max}$-$t_{min}$~100s. **e**, Separation distance versus time for the trajectory indicated by the micrographs in (**d**). **f**, Potential and force versus distance corresponding to (**e**). Red curves in (**b**) and (**e**) are the best fits of experimental data with $r_c(t)=(r_0^3-3\alpha t)^{1/3}$ for $r_0$=16.7 μm, $\alpha$=19.7 μm³ s⁻¹ in (**b**) and $r_0$=14.4 μm, $\alpha$=-15.8 μm³ s⁻¹ in (**e**). Blue curves in (**c**) and (**f**) are the best fits to a potential function proportional to $1/r_c$. Elastic charges estimated from fitting are 0.49 μm in (**b**) and 0.44 μm in (**e**). Micrographs in (**a**) and (**d**) insets are scaled such that their dimensions match the coordinate axes.

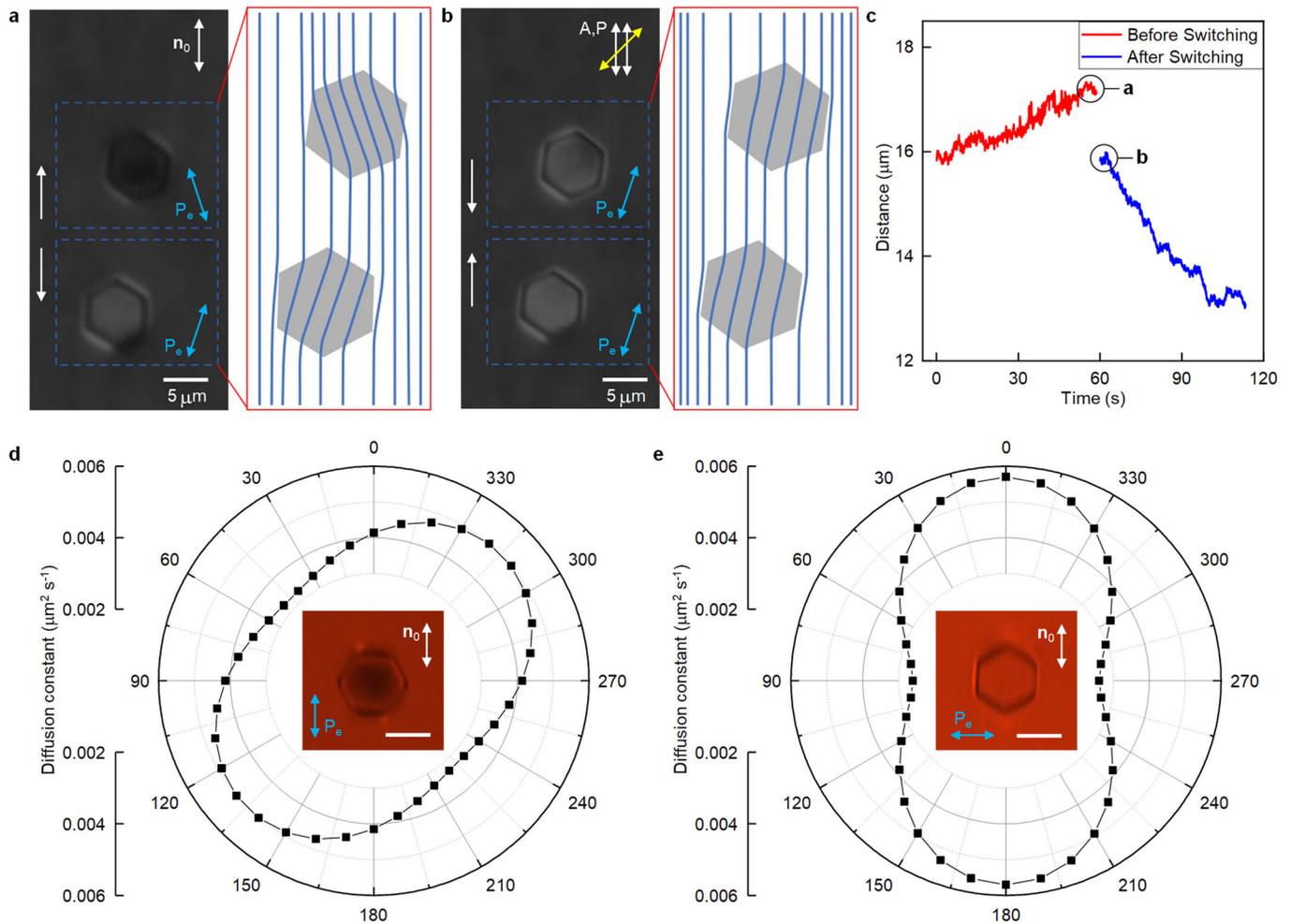

**Fig. 3 | Optical switching of the elastic monopole signs and interactions. a**, Micrograph (left) and corresponding schematics (right) of **n(r)** showing elastic monopoles of opposite sign induced by exposure to linearly polarized blue light at different polarizations. Regions exposed to differently polarized blue light are shown in dashed boxes, with the linear polarization directions marked by blue double arrows. Interaction forces are shown with a pair of white arrows. **b**, Micrograph (left) and corresponding schematics (right) of **n(r)** showing elastic monopoles of the same sign switched from the initial state shown in (**a**) by changing polarization of blue light to be the same for both platelets. The sign of monopole of the upper particle is then flipped and so is the direction of the interaction force. **c**, Separation distance versus time before and after the switching at the elapsed time of 58 s since the time when platelets were released from the laser traps. Points circled and marked "**a**" and "**b**" correspond to the micrographs shown in (**a**) and (**b**). **d,e**, Angular dependencies of diffusion constant of a platelet at $\mathbf{P}_e \parallel \mathbf{n}_0$ (**d**) and $\mathbf{P}_e \perp \mathbf{n}_0$ (**e**); scale bars in the insets

are 5 μm.

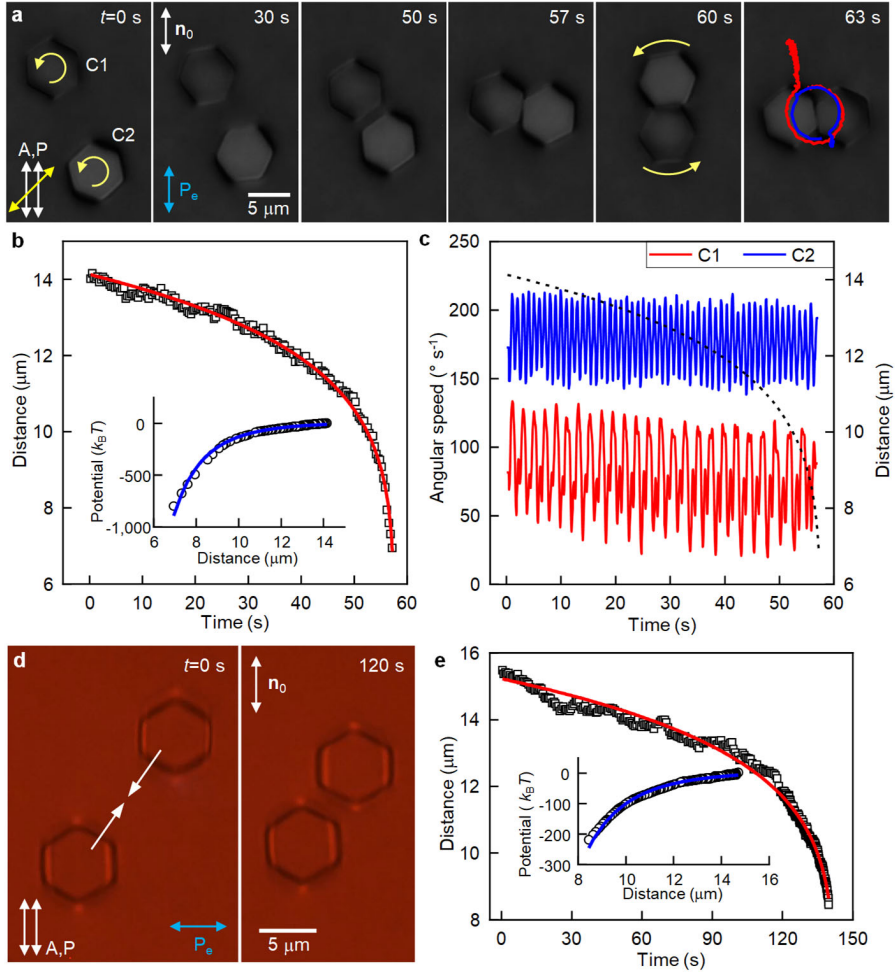

**Fig. 4 | Out-of-equilibrium colloidal assembly. a**, Elapsed-time-stamped micrographs showing two colloidal platelets (C1 and C2) attracting and assembling while spinning with $\mathbf{P}_e \parallel \mathbf{n}_0$. Red and blue lines overlaid on the last micrograph are trajectories of the two particles (red for C1 and blue for C2). **b**, Separation distance versus time of attracting particles till the moment of contact corresponding to (**a**). Inset shows interaction potential versus distance. **c**, Angular speed of the two spinning particles in (**a**), both exhibiting periodic motion, but at different frequencies. Black dash line indicates the distance between C1 and C2 versus time extracted from (**b**). **d**, Micrographs for different positions of platelets attracting diagonally with respect to $\mathbf{n}_0$ at $\mathbf{P}_e \perp \mathbf{n}_0$. **e**, Separation distance versus time corresponding to (**d**), with inset showing interaction potential versus distance. Red curves in (**b**) and (**e**) are the best fits of experimental data with $r_c(t)=(r_0^7-7\alpha t)^{1/7}$, $r_0=14.1$ μm, $\alpha=2.79\times10^5$ μm$^7$ s$^{-1}$ in (**b**) and $r_0=15.2$ μm, $\alpha=1.91\times10^5$ μm$^7$ s$^{-1}$ in (**e**). Blue curves in (**b**) and (**e**) are the best fits with $\propto -1/r_c^5$.

**Extended Data Figures**

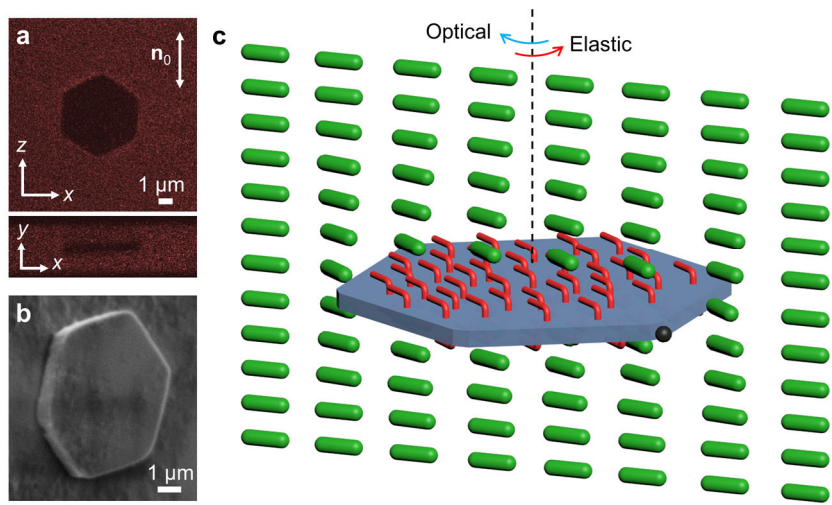

**Extended Data Fig. 1 | Hexagonal colloidal platelet with azobenzene monolayers dispersed in a nematic liquid crystal. a**, Three-photon excitation fluorescence microscopy images of a platelet in a LC cell obtained for the cross-sectional plane (*xz* plane) passing through the platelet's middle and for the vertical plane (*xy* plane) that is orthogonal to the plane of cell and the large area faces of the platelet. Both planes pass through the center of mass of the particle. **b**, Scanning electron microscopy image of an individual platelet placed on a substrate. **c**, Schematic of a platelet suspended in a nematic LC. Green rods indicate the director field **n**(**r**); red surface-attached rods indicate the orientation of the azobenzene molecules on the surface of the platelet; black semi-sphere on one of the vertices of the hexagon represents the surface point defects called "boojums". The optical torque that rotates the platelet away from its equilibrium state is balanced by the counteracting elastic torque due to director twisting, as shown by the blue and red curved arrows.

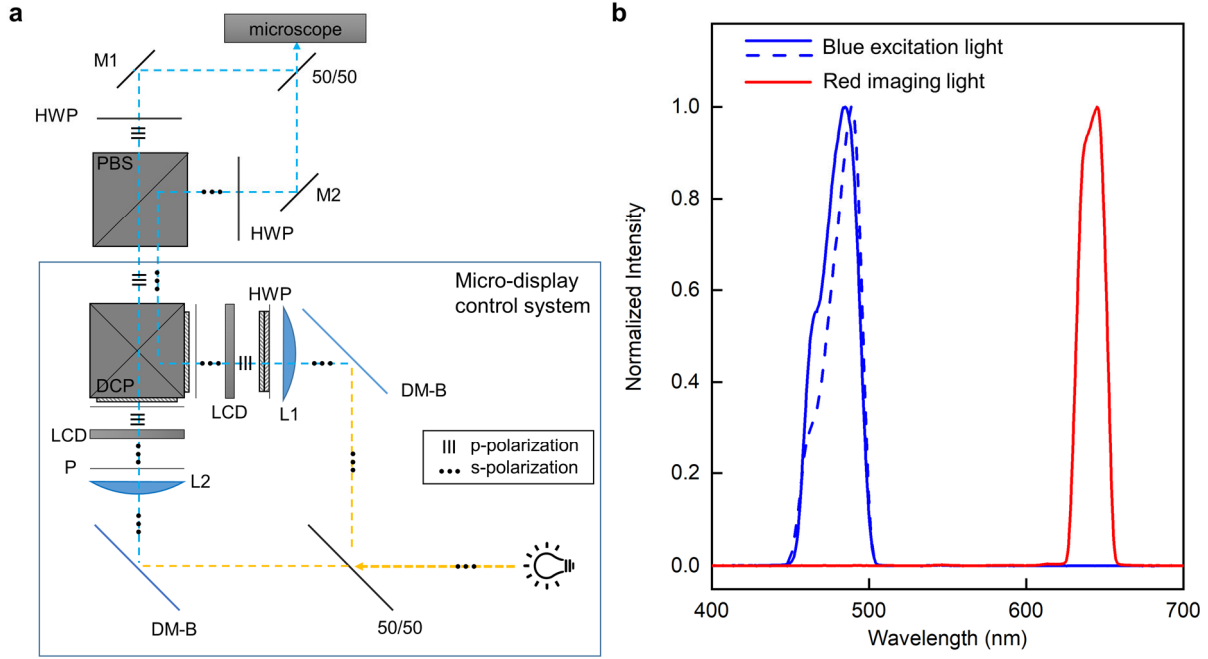

**Extended Data Fig. 2 | Micro-display-based illumination control system setup. a**, Schematic of the optical illumination setup. White light from the lamp source is split in half and directed into two optical paths using a beam splitter, filtered by two separate dichroic mirrors to supply blue light to two of the LC microdisplays (LCDs). After passing through the microdisplays, which define the spatial patterns of illumination, the two beams of light are recombined by a dichroic prism and then separated again based on their polarization. The two separated light paths correspond to patterns generated on the two LCDs and their polarization is further controlled by the half-wave plates inserted. Additional mirrors allow for fine-tuning the position of projected patterns after the light is re-combined and coupled into the microscope. In the schematic, 50/50s are plate beamsplitters (BSW10R, Thorlabs); DM-Bs are dichroic mirrors that reflect blue light; L1 and L2 are convex lens; HWPs are half-wave plates; Ps are polarizers; DCP is a dichroic cross prism; PBS is a cube polarizing beam splitter (CCM1-PB251, Thorlabs); M1 and M2 are silver mirrors. **b**, Spectra of the red imaging light from filtering light of the microscope lamp and the blue excitation light from the micro-display control system. Solid and dashed blue lines are the spectra of the two blue channels when turned on separately.

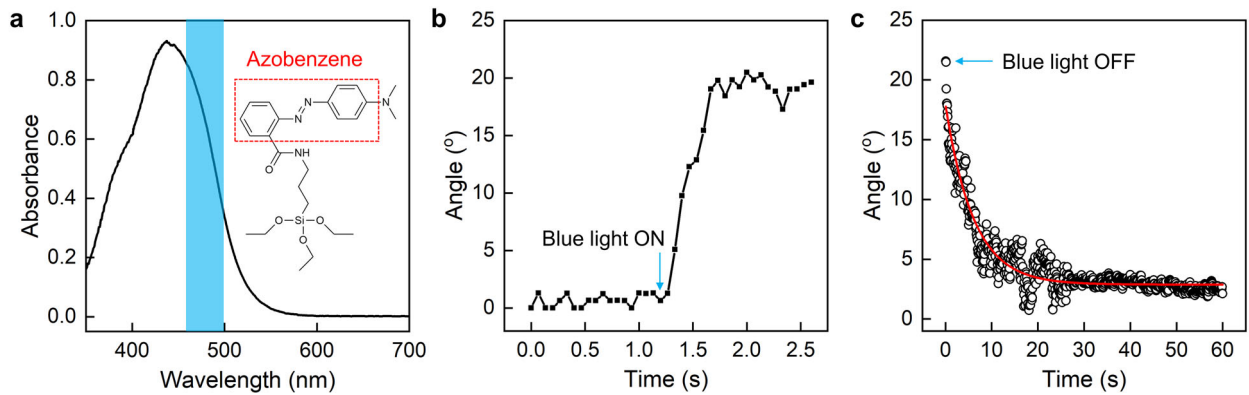

**Extended Data Fig. 3 | Optical response of the azobenzene-containing molecules and monopole-like platelets in a nematic LC. a**, Absorbance spectrum of the used photosensitive molecules of derivative methyl red in toluene at concentration of $5 \times 10^{-5}$ M. The spectrum was obtained using a 1-cm-thick cuvette with a spectrometer (Cary 500, Varian). **b**, Time dependence of the azimuthal orientation angle of the platelet upon switching on the blue excitation light at time $t \approx 1.2$ s, showing how the monopole moment is turned on upon light exposure. **c**, Time dependence of the azimuthal orientation angle of the platelet relative to its equilibrium position upon switching off the blue excitation light at $t=0$ s.

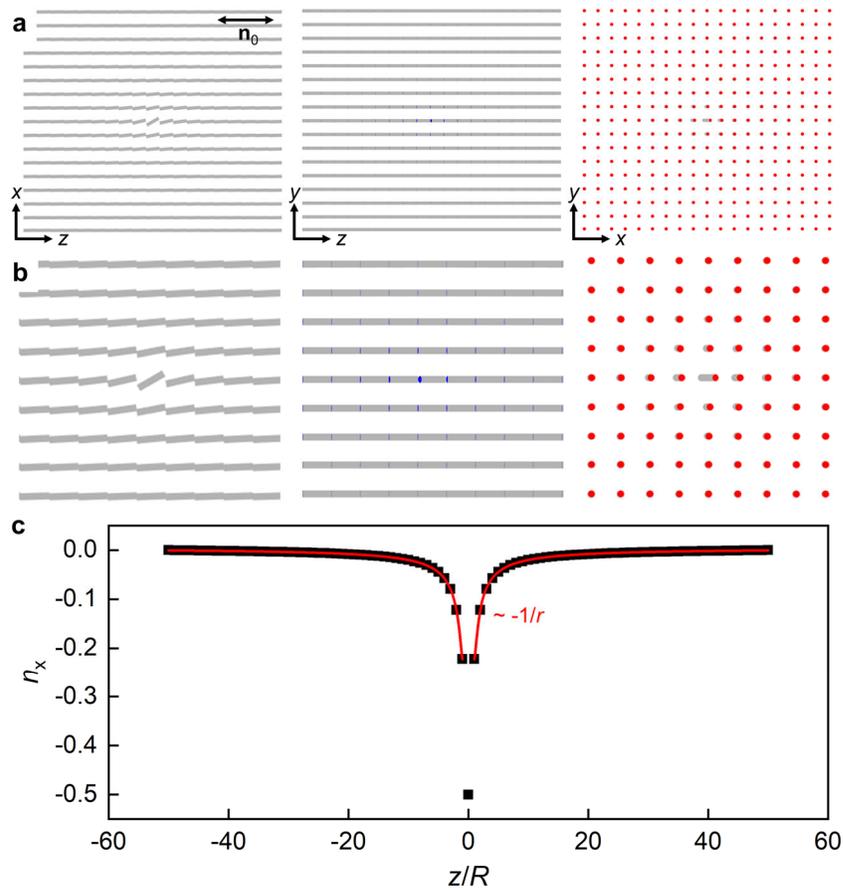

**Extended Data Fig. 4 | Computer simulated director configuration of an elastic monopole.**
**a**, Cross-sections of the director structure in *xz*, *yz* and *xy* planes passing through the center of the structure. Local director orientation is shown by grey cylinders with blue and red ends. **b**, Zoomed-in view of the cross sections in (**a**). **c**, Distance dependence of director deviation $n_x$ along *z* axis. Simulated results (black dots) are fitted by the anticipated for monopoles function $n_x \propto -1/r$ (red lines). *R* represents the effective particle size.

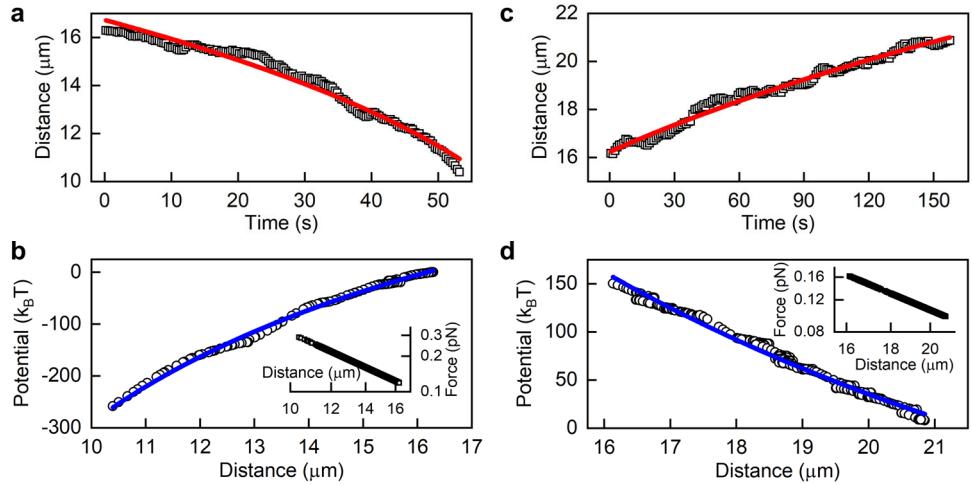

**Extended Data Fig. 5 | Characterization of interactions between elastic monopoles**. **a,c**, Separation distance versus time of monopolar attraction (**a**) and repulsion (**c**). Red lines in (**a**) and (**c**) are the best fits of the experimental data with the function $r_c(t)=(r_0^3-3\alpha t)^{1/3}$ and fitting coefficients $r_0$=16.7 μm, $\alpha$=21.1 μm$^3$ s$^{-1}$ in (**a**) and $r_0$=16.2 μm, $\alpha$=-10.6 μm$^3$ s$^{-1}$ in (**c**). **b,d**, corresponding potential and force (insets) dependencies on distance. Blue lines are the best fits of the experimental data with the function $\propto \pm 1/r_c$. The elastic charge estimated from the fitting parameters is 0.51 μm in (**a**), (**b**) and 0.36 μm in (**c**), (**d**).

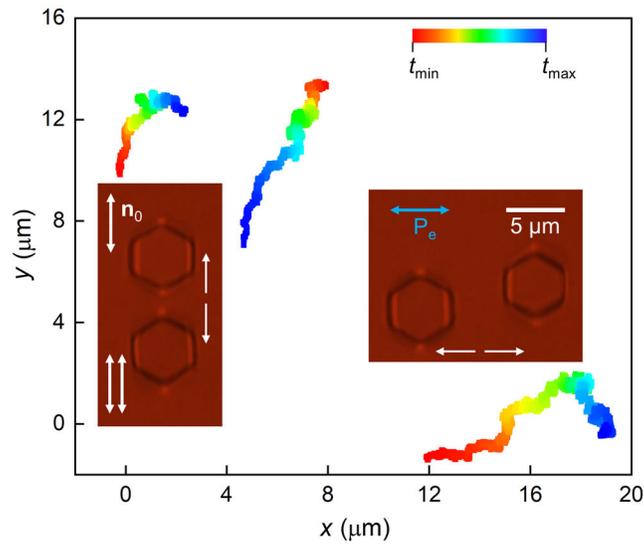

**Extended Data Fig. 6 | Angular dependence of quadrupolar interactions.** Trajectories are color coded with elapsed time and the duration of interaction $t_{max}$-$t_{min}$ is ~100 s. Insets are micrographs of platelets repelling each other when the separation vector that points from the center of one particle to that of the other is roughly parallel or perpendicular to $\mathbf{n}_0$. Images are taken under parallel polarizers shown in white double arrows. In contrast, the platelets attract when the separation vector is at an angle with respect to $\mathbf{n}_0$, as shown by the trajectory in the middle; the corresponding distance versus time dependence is shown in Fig. 4e.

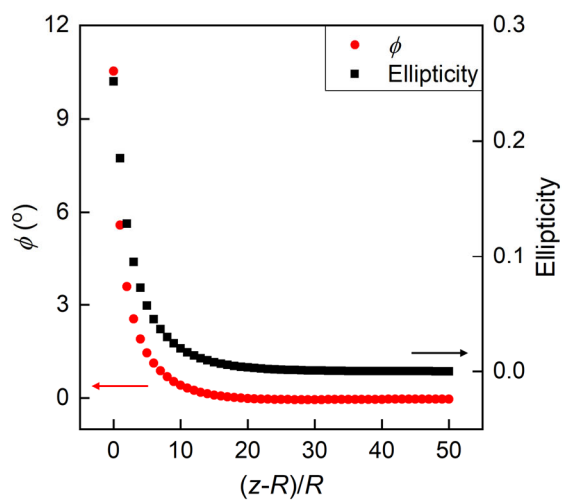

**Extended Data Fig. 7 | Computer simulated polarization states of the red imaging light after passing through the sample.** The angle $\phi$ between the long axis of the polarization ellipse and $\mathbf{n}_0$ as well as the corresponding ellipticity are shown with red dots and back boxes. Horizontal axis $(z-R)/R$ represent the distance from the edge of the particle along $z$ direction relative to particle radius $R$.

**Supplementary video legends**

**Video 1 Optical switching of the elastic monopole signs and interactions.** Elastic monopoles of opposite signs are first induced by exposure to linearly blue light beams with different linear polarizations. Then, changing the polarization of the blue light that excites the upper particle flips the sign of the induced monopole to be the same as that of the lower particle. Following this, the direction of the interaction also switches from repulsion to attraction. "P" and "A" mark parallel polarizer and analyzer of the microscope with a quarter-wave plate inserted in-between (fast axis of the waveplate is shown with an orange double arrow). Regions exposed to polarized blue light are shown in dashed boxes, with the corresponding linear polarization directions marked with blue double arrows. The direction of interaction force is shown with a pair of white arrows on the left.

**Video 2 Out-of-equilibrium colloidal interaction.** Two colloidal platelets attract while spinning and later assemble while continuing to spin in the same direction. Polarization of the blue excitation light (blue double arrow) is parallel to $\mathbf{n}_0$.